\input{aipcheck}

\documentclass[
    ,final            
  ]
  {aipproc}

\layoutstyle{6x9}

\usepackage{amssymb,amsmath}
\usepackage{wrapfig}
\newcommand{\captionsize}{\fontsize{9}{10}\selectfont}

\begin{document}

\title{Mixed Axion/Axino Dark Matter in mSUGRA and Yukawa-unified SUSY}

\classification{11.30.Pb, 12.60.Jv, 14.80.Mz}
\keywords      {Supersymmetric Phenomenology, Axion Dark Matter, mSUGRA, Yukawa-unified SUSY}

\author{Heaya Ann Summy}{
  address={Department of Physics and Astronomy, University of Oklahoma, Norman, OK 73019, USA}
}



\begin{abstract}

  Axion/axino dark matter (DM) is explored in the minimal supergravity (mSUGRA) and Yukawa-unified supersymmetric grand-unified theory (SUSY GUT) models with surprising results.  For this type of scenario, relic DM abundance has three components: {\it i}.) cold axions, {\it ii.}) warm axinos from neutralino decay, and {\it iii.}) cold or warm thermally produced axinos.  Reheat temperatures $T_R$ exceeding $10^6$ GeV are required in order to solve the gravitino/Big Bang Nucleosynthesis (BBN) problem while also allowing for baryogensis via non-thermal leptogenesis.  In order to attain high enough reheat temperatures, we also need high values of the Peccei-Quinn (PQ) breaking scale $f_a$ on the order $10^{11}$-$10^{12}$ GeV.

\end{abstract}

\maketitle


\section{Constraints on Axion/Axino Dark Matter}
\subsection{The Gravitino Problem}
In supergravity (SUGRA) models, both neutralinos and gravitinos are plagued by the gravitino problem.  Thermally-produced gravitinos $\tilde{G}$ too weakly coupled to be in thermal equilibrium can decay into various sparticle-particle pairs with long lifetimes of order $1$-$10^5$ seconds due to the Planck mass suppression of the gravitino coupling constant: $m_{\tilde{G}}=F/(\sqrt{3}M_{\mathrm{P}})$.  The energy injection during or after BBN from these late-decaying gravitinos can dis-associate light element nuclei created during BBN, thereby disrupting the successful predictions of the light element abundances calculated from nuclear thermodynamics.  
One way to avoid over-production of gravitinos in the early universe, and thus bypass the gravitino problem, is to require $T_R \lesssim 10^5$ GeV for $m_{\tilde{G}} \lesssim 5$ TeV.  However, such low values of $T_R$ rule out many attractive baryogenesis mechanisms, so we instead relax the upper bound to $T_R \lesssim 10^9$ GeV by allowing $m_{\tilde{G}} \gtrsim 5$ TeV.  As long as we keep $T_R$ below the upper bound, we can still avoid the gravitino problem.

\subsection{Non-thermal Leptogenesis}
The matter-antimatter asymmetry of the universe provides us with an additional cosmological constraint.  
%
In light of evidence for massive neutrinos, the currently popular mechanism for baryogensis is leptogenesis.  In thermal leptogenesis, heavy right-handed neutrino states $N_i$ ($i=1$-3 is a generation index) decay asymmetrically to leptons over anti-leptons in the early universe.  The lepton asymmetry is converted to baryon asymmetry via sphaleron effects.  The measured baryon abundance can be met for $T_R > 10^9$ GeV~\cite{thmlepto} subjecting this mechanism to the gravitino problem described above.  If we instead turn to non-thermal leptogenesis, where the $N_i$ states may be produced via inflaton decays, then we can achieve lower reheat temperatures.  The Boltzmann equations for the $B$-$L$ asymmetry (which is again converted into baryon asymmetry by sphaleron processes) have been solved numerically in Ref.~\cite{nonthmlepto}, within which the baryon-to-entropy ratio is found to be
\begin{equation}
\frac{n_B}{s}\simeq 8.2\times 10^{11}\times \left ( \frac{T_R}{10^6\;\mathrm{GeV}}\right ) \left ( \frac{2M_{N_1}}{m_{\varphi}}\right ) \left ( \frac{m_{\nu_3}}{0.05\;\mathrm{eV}} \right ) \delta_{\mathrm{eff}},
\end{equation}
where $m_{\nu_3}$ is the heaviest (active) neutrino mass, $M_{N_1}$ is the mass of the lightest right-handed neutrino, and $\delta_{\mathrm{eff}}$ is an effective $CP$-violating phase.  Comparing this to the WMAP value of $0.9\times 10^{-10}$~\cite{WMAP:baryonent} constrains the lower bound on the reheat temperature to $T_R \gtrsim 10^6$ GeV.  
%
A third type known as Affleck-Dine leptogenesis is described succinctly in Ref.~\cite{adlepto}.  The baryon-to-entropy ratio for Affleck-Dine can be expressed as
\begin{equation}
\frac{n_B}{s}\simeq \frac{1}{23}\frac{\left | \left < H \right > \right | ^2 T_R}{m_\nu M_{\mathrm{Pl}}^2},
\end{equation}
where $\left < H \right >$ is the Higgs field vev, $m_\nu$ is the mass of the lightest neutrino, and $M_{\mathrm{Pl}}$ is the Planck scale.  From this relation, values of $T_R \sim 10^6$-$10^8$ GeV are allowed for $m_\nu \sim 10^9$-$10^7$ eV.

\section{Axion and Axino Production}
The lifetime of axions is longer than the age of the universe (their decay mode is $a\rightarrow \gamma \gamma$), so they can be a good dark matter candidate.  For reheat temperatures $T_R \lesssim 10^9$ GeV $< f_a$, only axion production via vacuum misalignment~\cite{vacalign} is relevant.  Then the axion field $a(x)$ can start with any value on the order of the Peccei-Quinn (PQ) breaking scale at temperatures $T\gg\Lambda_{\mathrm{QCD}}$.  As the temperature of the universe drops, the potential turns on causing $a(x)$ to oscillate and settle to its minimum at $-\bar{\theta}f_a/N$ (here $\bar{\theta}=\theta + \mathrm{arg}[\mathrm{det}\; m_q]$, $\theta$ is the fundamental strong {\it CP}-violating Lagrangian parameter, and $m_q$ is the quark mass matrix).  The difference between $a(x)$ before and after the potential turns on corresponds to the vacuum misalignment producing an axion number density
\begin{equation}
n_a(t)\sim \frac{1}{2}m_a(t)\left< (a^2(t)\right>,
\end{equation}
where {\it t} is the time near the QCD phase transition.  Relating the number density to the entropy density yields a formula for the axion relic density today~\cite{vacalign}:
\begin{equation}
\Omega_ah^2 \simeq \frac{1}{4}\left(\frac{6\times 10^{-6}\;\mathrm{eV}}{m_a}\right)^{7/6}.
\end{equation}
Axions produced from vacuum misalignment constitute {\it cold} dark matter (CDM).

Axinos can be produced thermally or non-thermally.  If the axino $\tilde{a}$ is the lightest SUSY particle (LSP), then the neutralino can decay via $\tilde{\chi}^0_1\rightarrow \tilde{a}\gamma$.  The axinos inherit the thermally produced neutralino number density.  Thus, the non-thermal production (NTP) relic abundance of axinos is
\begin{equation}
\Omega_{\tilde{a}}^{\mathrm{NTP}}h^2 = \frac{m_{\tilde{a}}}{m_{\tilde{\chi}^0_1}}\Omega_{\tilde{\chi}^0_1}h^2.
\end{equation}
NTP axinos are {\it warm} dark matter for $m_{\tilde{a}}<1$-10 GeV~\cite{NTPaxinos}.  The neutralino-to-axino decay can reduce relic density by large factors, thereby reinstating models that otherwise yield too much dark matter.

Axinos can be thermally produced via scattering and decay processes in the cosmic soup so that they need not be in thermal equilibrium.  Using hard thermal loop resummation, the axino thermally produced (TP) relic abundance is given in Ref.~\cite{BS} as
\begin{equation}
\Omega_{\tilde{a}}^{\mathrm{TP}}h^2\simeq 5.5g_s^6\ln\left(\frac{1.211}{g_s}\right) \left(\frac{10^{11}\;\mathrm{GeV}}{f_a/N}\right)^2 \left(\frac{m_a}{0.1\;\mathrm{GeV}}\right) \left(\frac{T_R}{10^4\;\mathrm{GeV}}\right),
\end{equation}
where $g_s$ is the strong coupling evaluated at $Q=T_R$ and $N(\sim \mathcal{O}(1))$ is the model-dependent color anomaly of the PQ symmetry.  Thermally produced axinos with mass at least 100 keV qualify as {\it cold} dark matter.

\section{Mostly Axion Cold Dark Matter in $\mathrm{\bf m}$SUGRA}
\begin{figure}
  \begin{minipage}{.5\textwidth}
    \includegraphics[height=.2\textheight]{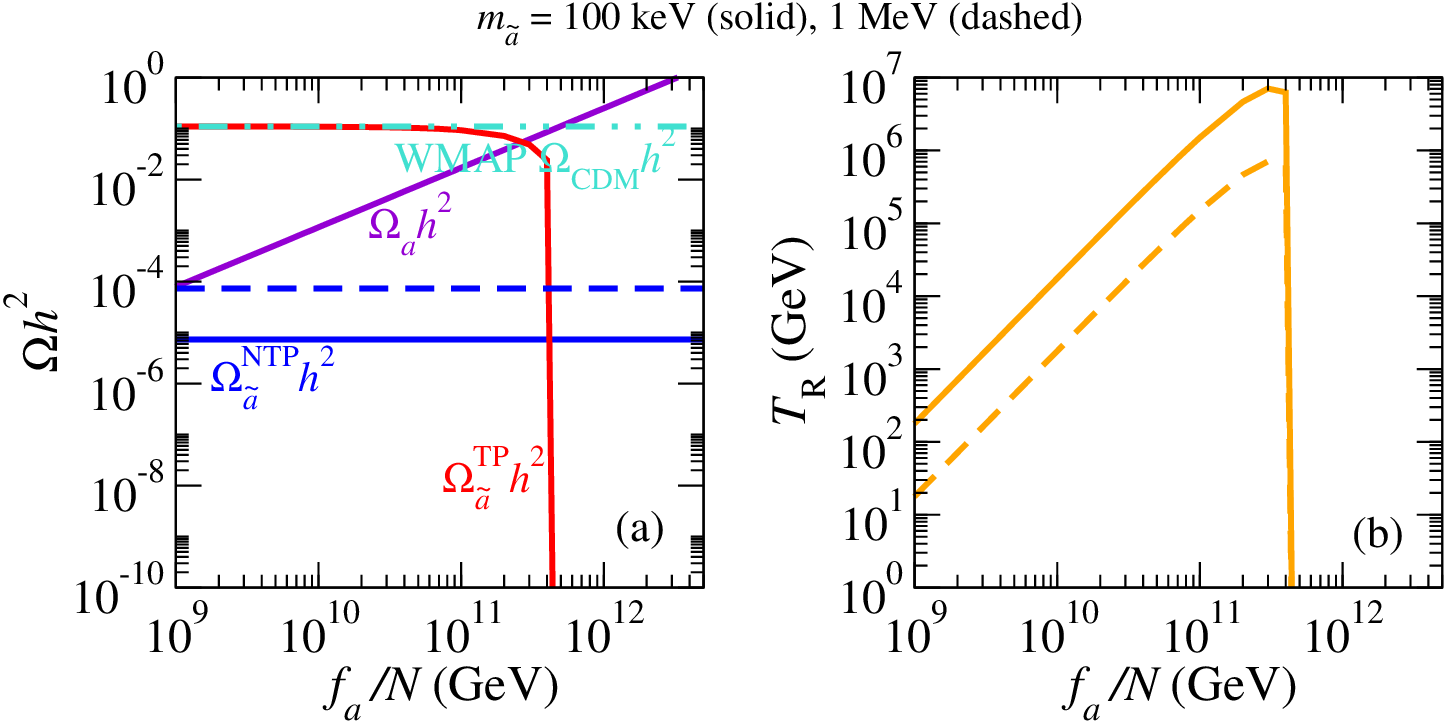}
  \end{minipage}
  \hskip .12\textwidth
  \begin{minipage}{.38\textwidth}
    {\captionsize {\bf FIGURE 1.}$\;\;\;\;$(a) Axion, TP axino, and NTP axino contributions to dark matter density for the two $m_{\tilde{a}}$ values noted above the figure as a function of the PQ scale.  (b) Similarly, $T_R$ is plotted.}
  \end{minipage}
\end{figure}
Fig.~1 shows results for a three-component combination of axions, TP axinos, and NTP axinos for mSUGRA point with $(m_0$, $m_{1/2}$, $A_0$, $\tan\beta$, $\mathrm{sgn}[\mu])=(1000,300,0,10,+1)$ plus $m_t=172.6$, with all mass parameters in units of GeV.  Clearly, values of $f_a/N\gtrsim 10^{11}$ GeV needed to surpass the lower bound on $T_R\; (\gtrsim 10^6$ GeV$)$ increasingly favor mostly axion dark matter.

A side-by-side comparison of parameter space available in mSUGRA for (a) neutralino CDM and (b) axion CDM is shown in Fig.~2.  These slices through the $m_0$-$m_{1/2}$ plane are for $A_0=0,\;\tan\beta=10,$ and $\mu>0$.  Additionally, frame (b) is evaluated at $f_a/N=3\times 10^{11}$ GeV, $\Omega_ah^2=0.11$, $\Omega_{\tilde{a}}^{\mathrm{TP}}h^2=0.006$, and $\Omega_{\tilde{a}}^{\mathrm{NTP}}h^2=6\times 10^{-6}$ such that axino DM is used to fill the error bars from WMAP 
observations.  Axions and axinos have again made available much of the mSUGRA parameter space that has been ruled out under neutralino DM!
\setcounter{figure}{1}
\begin{figure}
  \begin{minipage}{.5\textwidth}
    \includegraphics[height=.22\textheight]{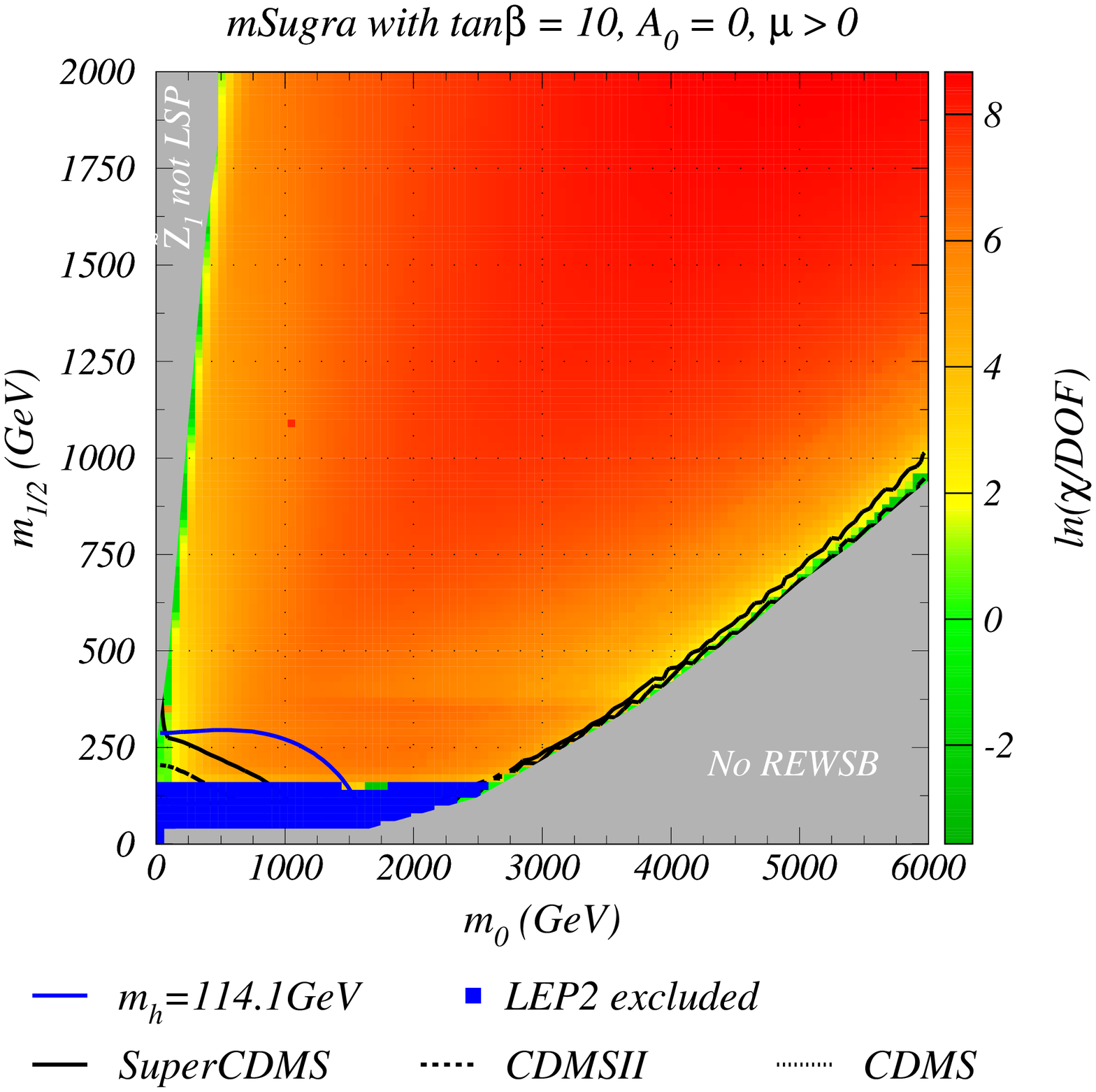}
  \end{minipage}
  \hspace{-.1\textwidth}
  \vspace{-.25\textheight}
  \begin{minipage}{.5\textwidth}
    \includegraphics[height=.22\textheight]{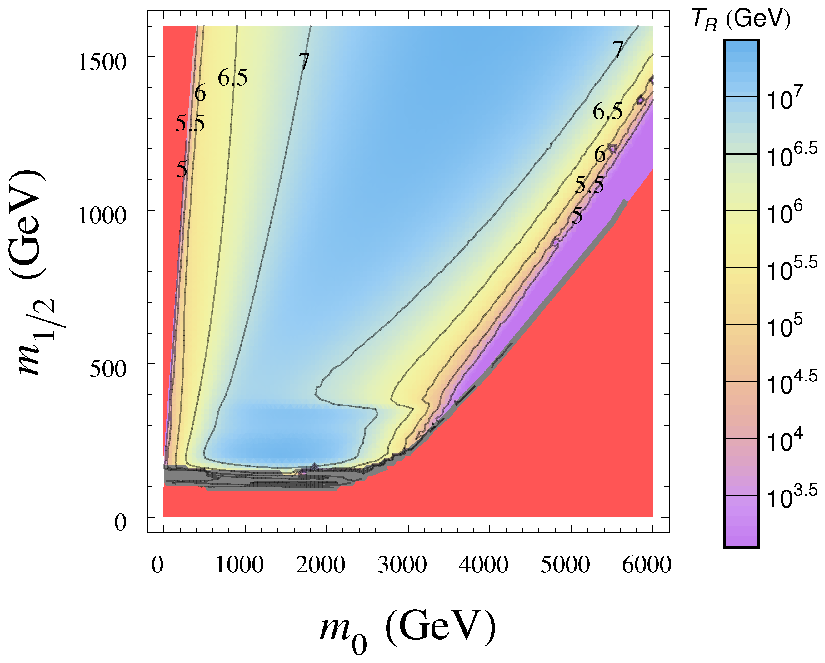}
  \end{minipage}
  \caption{(a) Allowed regions for neutralino DM (green) using a $\chi^2$-fit.  (b) Favorable axion DM regions (blue) for reheat temperatures above $\sim 10^6$ GeV.}
\end{figure}

\section{Mixed Axion/Axino DM in Yukawa-unified SUSY}
It was previously mentioned that the mechanism for producing NTP axinos can reduce large values of relic density.  One such scenario where this happens is in Yukawa-unified SUSY GUTS.  For example, a case study in Ref.~\cite{DMso10} (using the Higgs-splitting method to accommodate electroweak symmetry breaking minimization conditions) scanning over Yukawa-unified $SO(10)$ parameter space with $(m_{16}:0$-20 TeV, $m_{10/16}:0$-1.5$,0$-5 TeV, $A_0/m_{16}:-3$-3, $M_D/m_{16}:0$-0.8, $\tan\beta:40$-60$)$ yields neutralino relic abundance exceeding $\sim 10^3$.  By adoping axion/axino DM, Ref.~\cite{so10cosm} indicates that for Yukawa-unified models, again mostly axion CDM with perhaps a small axino component fits within cosmological requirements for $T_R$ and $\Omega_{\mathrm{CDM}}h^2$ (in addition, $f_a/N$ needs to be large and heavy scalars seem to be favored in these types of models).

\end{document}